\documentstyle[epsfig]{aipproc}
\font\gkvec=cmmib10                         
\def\bomega{\hbox{{\gkvec\char33}}}     
\def\ni{\bibitem{}}
\def\ts{\times}
\def\lb{\langle}
\def\rb{\rangle}
\def\curl{\nabla {\ts}}

\def\bbw{\overline{\bf \bomega}}
\def\bfwp{{\bf \bomega}'}
\def\bfvp{{\bf v}'}
\def\bfbp{{\bf b}'}
\def\bb0{\bfbp^{(0)}}
\def\bv0{\bfvp^{(0)}}
\def\bw0{\bfwp^{(0)}}
\def\gsim{\raise2.90pt\hbox{$\scriptstyle>$} \hspace{-6.4pt}
\lower.5pt\hbox{$\scriptscriptstyle
\sim$}\; }
\def\lsim{\raise2.90pt\hbox{$\scriptstyle<$} \hspace{-6pt}\lower.5pt\hbox{$\scriptscriptstyle\sim$}\; }

\begin{document}
\title{Some Consequences of Magnetic Fields in High Energy Sources}

\author{Eric G. Blackman}
\address{Department of Physics and Astronomy, University of Rochester, 
Rochester, NY 14627} 

(to appear in ``Proceedings of The First KIAS Astrophysics Workshop,
"Explosive Phenomena in Astrophysical Compact Objects", 2000,
edited by H.-Y. Chang, C.-H. Lee, M. Rho, I. Yi,
AIP Conference Proceedings)


\maketitle

\begin{abstract}
{\small Magnetic fields likely 
play a fundamental intermediary role between gravity
and radiation in many astrophysical rotators. They can, among other
things, 1) induce and be amplified by turbulence, 
2) energize coronae, 3) launch and collimate 
outflows in ``spring'' or ``fling'' mechanisms  
The first is widely recognized to be important for angular momentum transport, 
but can also produce intrinsic variability and vorticity growth.  
The second leads to the production
of high energy flares, and also facilitates a test of general relativity
from AGN observations. The third can operate from 
rotators with a large scale fields,  
though the origin of the requisite large scale fields is somewhat unresolved.
I discuss these three points in more detail below, 
emphasizing some open questions.} 
\end{abstract}

\section*{1. Turbulence, Variability \& Vorticity in Accretion Disks}
%
%
%
%

Accretion disks are a widely accepted paradigm 
to explain  a variety of spectral features in 
sources such as active galactic nuclei (AGN), X-ray binaries, 
cataclysmic variables (CVs), dwarf novae and protostellar 
[1,2,3].  As gas orbits a central massive source, internal dissipation  
drains the orbital energy, allowing material to move in and
angular momentum out.  Some 
fraction of the dissipated energy accounts for  the observed luminosity.  
Micro-physical viscosities are often 
too inefficient 
so an enhanced transport mechanism, likely involving turbulence, essential.
The observational evidence for an enhanced/turbulent viscosity is 
least direct in the case of  AGN disks,
though it is natural to expect that the latter 
would be subject to the same turbulent viscosity generating mechanisms:
As astrophysical disks are likely seeded with a magnetic field, 
the ``Balbus-Hawley'' or magneto-rotational instability (MRI) [3,4]
ensues.  This produces self-sustaining turbulence
for flows with a radially decreasing angular speed that transports
angular momentum in rotationally supported disks.  
Turbulence driven by the MRI transports angular momentum outward,  
unlike convection [3,5,6]. 
Rossby wave vortices may also transport angular momentum outward [7].

For low enough accretion rates, the dissipated energy may be
primarily advected by hot protons in advection dominated accretion 
flows (ADAFs) rather than radiated, possibly explaining why some sources 
have engines that are surprisingly quiescent [9,10,11].
For optically thin, geometrically thick ADAFs, the 
disks may be physically similar to the coronae above
the thin disks of Seyferts [12].
The thin disk+corona is something like a 
hamburger sandwich with the optically thin (geometrically thick) 
corona being the bun and the optically thick
(geometrically thin) disk is the meat.  The 
ADAF is a sandwich with no ``meat.''   

Alternatives to simple ADAFs suggest that  
quiescent luminosities might instead result from reducing the
accretion rates into the inner regions [13,14,15].
Such models include advection dominated inflow outflow solutions (ADIOS)
[13] and  convection dominated accretion 
flows (CDAFs) ([14]; Narayan, these proc).
These require a less than maximal viscosity [16].

\subsection*{Mean Field Disk Theory and Variability}

In general, the ubiquity of turbulence in disks has important
implications for what the accretion disk equations mean.  
Analytic disk equations are mean field equations.
While non-linear instabilities in thin disks 
have been extensively simulated locally 
[17,18] a traditionally useful approach to disk 
models has been to swipe the details of the stress tensor
into a turbulent viscosity of the form [19]
\begin{equation}
\nu_{tb}=\alpha c_s H\simeq v_{tb}l_{tb},
\label{1}
\end{equation}
where $H$ is the disk height, $c_s$ is the 
sound speed, $l_{tb}$ is the dominant turbulent scale, 
$v_{tb}$ is the eddy speed at that scale, 
and $\alpha_{ss}< 1$ is a constant.  
Use of this formalism 
requires a mean field theory, not a replacement of the
molecular viscosity with the turbulent
viscosity.  Assumptions of azimuthal symmetry
and steady radial inflow [1,11] 
require turbulent motions to be averaged 
over the time and/or spatial scales on which mean quantities vary.

The required mean field approach is complementary to
that employed in mean field magnetic dynamo theory,  
where the field is split into mean and fluctuating
components, ${\bf B}={\overline {\bf B}}+ {\bf B}'$, and the 
induction equation is averaged and solved.
For the accretion disk case, the momentum equation
must be similarly split, and the evolution equation for the 
mean velocity field derived. 
The result is that $\nu_{tb}$ 
represents a correlation of turbulent
fluctuations, that is 
$\nu_{tb} ={\lb {\bf v}'(t)\cdot \int {\bf v'}(t')dt'\rb}$
[20,21,22].

Mean field theory 
has a limited precision, and thus predicts variability.   
Let the mean represent a global average
over azimuth and half thickness, without  averaging over the
radius.  Since all mean velocities are scalings of the Keplerian
velocity, each radius is in principle ``labelled'' by its Keplerian speed.
However, the averages over fluctuations produce an uncertainty
in this labelling.  Since the luminosity at a given 
frequency depends on the radius of emission for accretion disk models,
we can relate the uncertainty in the observed luminosity
to the uncertainty in the radius. That is, 
\begin{equation}
\Delta L_{\nu} / L_{\nu} \simeq |R\partial_R({\rm Ln}L_\nu)| {{\Delta R}/ R}=\Psi {\Delta R}/ R
\label{var}
\end{equation}
where $\Psi$ can depend on temperature but $\sim 1$ 
for a range of frequencies for thin
and thick disk models [23].
Let us estimate ${{\Delta R}/ R}$. The ``error'' associated  with a single  
turbulent fluctuation of scale $l_{tb}$ 
is reduced  by $N^{1/2}$ for each 
averaged dimension,  where $N$ is the number of eddy spatial 
scales  averaged  over in that dimension.  As observational data are taken
in a time averaged sense, there is also a reduction that depends on 
observation time, $t_{obs}$.  The result is   
$\Delta R \simeq l_{tb}/[N_{\phi}N_z(1 + t_{obs}/t_{tb})]^{1/2}$,
where $N_\phi$ and $N_z$ are the number of dominant eddy scales
in the $\phi$ and $z$ directions,  
and $t_{tb}$ is the dominant correlation time.  

To make further progress, note that for fully developed MHD
turbulence, near equipartition between kinetic and magnetic energy
is generically reached, and is 
$v_{A}\sim 2^{1/2}v_{tb}$ 
[3,17,18] in the case of accretion disks. The 
factor of $\sqrt 2$ comes from shear that adds a bit more amplification
to the field. The MRI instability onset time scale 
is of order the dominant eddy turnover or cascade time and is 
$t_{tb}\sim l_{tb}/v_{tb}\sim \Omega^{-1}$, where $\Omega$ is the 
rotation speed. Using these and (1) we have 
\begin{equation}
v_{A}^2/2\Omega =\alpha_{ss}c_s H.
\label{ad0}
\end{equation}
Using 
$\Omega H\sim c_s$ for vertical pressure support, we then have 
\begin{equation}
\alpha_{ss}\simeq  v_A^2/2c_s^2\simeq (l_{tb}/H)^{2}.
\label{ad01}
\end{equation}
The next step is to note that $N_{\phi}=2\pi R/(2l_{tb})=
(\pi R/H\alpha_{ss}^{1/2})$,
where the extra 2 on the bottom comes from  eddy elongation
in the $\phi$ direction from shear, and 
$N_{z}=H/2l_{tb}=\alpha_{ss}^{-1/2}$, where the 2 is from the 1/2 thickness. 

Collecting all of the above into (2) gives 
\begin{equation}
\Delta L_{\nu}/L_{\nu}\sim 
\Psi \alpha_{ss}(H/R)^{3/2}
(1 + \Omega t_{obs})^{-1/2}. 
\label{1a}
\end{equation}
For thick disk ADAF models, 
$H/R \sim 1$ and $\alpha_{ss}\sim 1$ [8],
large variability around the predicted luminosities can be 
expected unless $t_{obs} >> \Omega^{-1}$.
When variability is not seen, or
a systematic deviation from the theory
is seen in a sample of observations at widely separated times or
in different objects, the simplest ADAF type model 
may not be capturing the physics. 
Such systematic deviations seem to be evident in some of the large
ellipticals 
[24] 
Other such quiescent accretor variations such as CDAFs
(Narayan these proc.) or ADIOS type models [13]
 which involve outflows may be more appropriate there.

\subsection*{Vorticity}

There are other effects of mean field theory in addition to the variability. 
If one considers the vertical disk averaging 
to be taken only over one hemisphere, 
then in addition to the scalar Shakura-Sunyaev turbulent viscosity 
transport term used in simple analytic accretion disk modeling, 
a pseudoscalar  transport term also arises [22].
This term is analogous to  that which appears in magnetic
dynamo theory, and can lead to vorticity growth [21,25].

In the same way that the mean field magnetic dynamo characterizes
an inverse cascade of magnetic helicity [26],
the vorticity dynamo highlights some growth of vorticity on 
larger scales than the input turbulence. 
Enstrophy exhibits an inverse cascade in 2-D turbulence
[27].  The growth of vorticity in
 primarily 2-D rotating fluids has been seen in nature (e.g. Jupiter [28])
 as well as in simulation [29] and experiment [30].  
Statistical mechanics approaches have modeled  this [31].  
Vorticity growth in sheared thin accretion disks has been studied less than 
th associated vortex evolution [32,33,34].

For an accretion disk in which mean quantities depend only
on radius, Ref. [22]  showed that 
\begin{equation}
\begin{array}{r}
d_{t}\bbw = \curl \alpha_{0} \bbw 
-\nabla\ts(\nu_{tb}\nabla\ts\bbw)
\end{array}
\label{WEQM}
\end{equation}
%
in the frame comoving with the cartesian velocity tangent
to the mean rotation, where the coefficients are
\begin{equation}
\begin{array}{l}
\alpha_{0} = (\tau_{c}/3)(\langle\bw0\cdot{\bfvp}^{(0)}\rb
\\
\nu_{tb} =(\tau_{c}/3)\langle\bfvp^{(0)}\cdot\bfvp^{(0)}+\bb0\cdot\bb0\rangle \\
\end{array}
\label{COEF}
\end{equation}
and $\bb0\equiv {\bf B}'^{(0)}/4\pi{\bar \rho}$, with $\bar \rho$
as the mean density. (This equation presumes that 1st order cross correlation
terms vanish, that is 
$\lb \bfbp^{(0)}\cdot\bfvp^{(0)}\rb
=\lb{\bfwp}^{(0)}\cdot{\bf b}^{'(0)}\rb=
\lb {\bfwp}^{(0)}\cdot\curl \bfbp^{(0)}\rb=0.
$)
Note that $\nu_{tb}$ is not the only transport term.
There is also the  pseudoscalar $\alpha_0$ term
as in the magnetic field case.  It is this pseudoscalar
term which can lead to vorticity growth. 
Interestingly, the pseudoscalar
for the vorticity, unlike for
that of the mean magnetic field, has a kinetic helicity term 
{\it without} a current helicty term [22].
The $\alpha_0$ can be parameterized  as $\alpha_0=q\alpha_{ss} c_s$.
One necessary condition for growth turns out to be $q>H/R$.

The simplest growth solutions [22]
show a dominant growth scale $\sim H$, leading to
intermediate scale vortices that should survive at least
a vertical diffusion time, that is, $\gsim$ few $\times\ 1/\alpha_{ss}$ orbits.
For $\alpha_{ss}\sim 0.01$,  
the resulting anti-cyclonic vortices may allow
dust trapping, catalyzing planet formation when applied
to star+planet system forming disks [31,33,35].  

Note that this simplified model of intermediate
scale vorticity growth cannot tell
how many vortices grow, or where in height or azimuth these
vortices are, only that there are  growing solutions.
This is because here the variables are averaged to depend only on radius.
Note also that if the vertical averaging is taken over the
full scale height, then the $\alpha_0$ coefficient should
vanish because the psuedoscalar reverses sign across
the mid-plane.  Then vorticity growth could not be
identified in this over simplified formalism.

\section*{2. Buoyancy, Coronae \& AGN Iron Line Profiles}

\subsection*{Coronae}

MHD turbulence likely involves  
spatial intermittency of the magnetic field 
[36], even in accretion disks.
While the dynamics of intermittency is not fully understood, 
the random component of the field would 
be preferentially amplified  at regions of strongest 
shear.  Since small scale shear varies spatially and spectrally 
in a turbulent medium, the field strength would also be expected to
to vary in correlation.   The strongest magnetic field regions might 
form a kind of dynamic sponge, intermixed with 
``void'' regions that have much weaker fields.
The extreme  situation, in which the magnetic
field occupies a distinct volume from the thermal material,
could maintain dynamical equilibrium when the ratio of the 
the average particle to magnetic pressure in the
disk ($\equiv \beta_p$) is large.   
To see this, consider the field to reside in magnetically dominated flux 
tubes.  The pressure inside balances the external 
particle pressure.  However, if $\beta_p >> 1$, 
the tubes would occupy only a small volume filling 
fraction ($\sim 1/{\beta_p}$) [37].  Since the distance between tubes 
would be large, intersections with other tubes would be too infrequent 
to significantly load particles into the tubes.  In contrast, if the magnetic 
volume filling fraction were large, frequent
reconnection events over a large
fraction of the tubes' longitudinal cross section would 
more easily mass load the tubes and lead to one phase medium.
The amount of intermittency remains to be understood.
Understanding this intermittency is important because
extremely evacuated tubes rise at speeds of order their
internal Alfv\'en speeds  and form coronae.
Coronal magnetic dissipation is as important a 
problem for disks [38] as it is for the Sun [39]

While coronae have been long thought to form above
disks [38,40] 
analytic work often does not incorporate the turbulence
and/or the origin of the initial large field.
The usually invoked Parker instability favors wave modes
which can be shredded by the Balbus-Hawley instability on time
scales approximately equal to rise times.
Simulations of turbulent disks
show that coronae do form in turbulent disks [41]
but the dominant mechanism, and its relationship to the process
of formation in the solar corona is not fully sorted out.

If the disk were not subject to  MRIs and turbulence,
the induction equation tells us that the disk field would 
grow linearly from shear.  It would saturate
at a value that is higher than the saturation value of the
MRI. This is because the shear can amplify the field over a much 
longer time in a laminar disk; the field remains coherent longer.
Field strengths of order the thermal pressure could incur before buoyancy
(in this case by the Parker instability) drained the field into the 
corona. For MRI driven disks, 
the shear  only operates on a single magnetic field filament for
a correlation time (one rotation) after which that filament loses its
identity.  The field energy saturates by a factor of $\alpha_{ss}$ lower
for the turbulent case. Also for the the laminar case, 
the corona would be more intermittent.
This is because  the buoyancy and the dissipation would likely
get rid of the field faster than it builds up.
For a laminar disk, the field growth would be linear in time 
whereas for the MRI disk,  the growth is exponential.  
A laminar disk + corona should thus exhibit
longer variability periods than a turbulent disk, but with larger
amplitudes.
The observational evidence 
in the case of AGN disks/corona seems to suggest 
a turbulent disk. While X-ray variabilities of factors of several 
are observed [42], factors of orders of magnitudes are not, 
suggesting a steady background level of coronal dissipation.

\subsection*{AGN Iron Lines and Engine Geometries}

Seyfert galaxy X-ray spectra, are best 
modeled as a combination of direct emission from a hot corona and  
reprocessed emission from a cold, optically thick accretion disk 
[43]. The direct component results from
inverse Compton scattering of thermal disk UV photons. The reprocessed
component incurs as photons are scattered back onto the disk.  

If it weren't for the active coronae, 
we would lose a probe of strong 
gravity [44]. The reprocessed coronal emission includes the 
broad iron K$\alpha$ fluorescence line of rest energy 6.4 keV,
which carries information about the geometry and dynamics of the reprocessing
material near the black hole [44]. 
ASCA has observed iron lines in $\sim 18$ Seyfert Is 
[45]. The best studied iron line is that of MCG-6-30-15 [46,47]. 
In addition to the geometry, the 
iron line profiles are sensitive to the disk illumination law, 
the disk inclination, and the inner and outer radii [44].
Most work on reprocessing in AGN has invoked thin  
flat disks with axially symmetric illumination laws 
representing a ``point'' X-ray source.

Some AGN line profiles  like MCG-6-30-15 are 
consistent with flat disks [46,47] 
but the current data may be not precise enough
to rule in or out non-axisymmetric engines in other cases.
Such engines are also worth considering in order to provide
robust comparisons to flat disk models. 
Ref [48] considered  finite disk thicknesses and 
Ref [49] considered concave disks.
Warped disks have also been considered around Schwarzchild black holes 
[50].  There are plausible theoretical reasons for disk warping  
including radiation driven warping [51] and  tidal warping [52].
On the observational side, water maser emission of NGC 4258 at  
0.1pc from the central engine (on larger scales than the inner accretion disk) 
traces a disk warp [53]. 
There may also be indirect evidence on these larger scales from
observations of Seyfert Is which 
suggest that the broad line regions are not coplanar with the inner disk [54]. 
Dusty tori of 
Seyfert unification paradigms might also involve  warped disks [55].

Warped disk studies reveal line features 
that are impossible for unobscured flat thin disks.
First, shadowing of the source by the disk 
and shadowing of reprocessed emission 
by the disk blocks regions of the disk from contributing to the iron line.  
Sharper red than blue cutoffs or very soft blue cutoffs can also arise.  
The latter characteristic is seen in some profiles of  Ref [45]. 
The sharp red cutoffs result for large inclination angles, which is 
consistent with some Seyfert IIs [56].
Second, non-axisymmetry of the disk means that line profiles can show 
time variability if the warp precesses around the disk.
Third, there can be sharper peaks near the rest 
frequency compared to a flat disk since concavity can 
offer more  solid angle covering fraction.
Fourth, apparent misalignment of central disk plane with the obscuring 
torus can be accounted for.

For a concave disk, sharper 
peaks near the rest frequency can be accompanied by a total
reprocessed emission fraction that is larger than 1/2 [49], 
where 1/2 is the maximum limit for a point source above a flat disk.
This may play a role in ultra-soft narrow-line 
Seyferts (Brandt, private communication 1999), though there
are other ways to achieve this enhanced reprocessing fraction.

Line profiles from a distribution of dense clouds in an optically thin,
geometrically thick disk may apply to ADAFs.
Dense clouds formed from thermal instability 
can survive long enough to produce reprocessing signatures in
otherwise optically thin flows [57].  This needs more
investigation.

\section*{3. Springs, Flings \& Large Scale Fields}

Many large scale jets and winds in astrophysics 
including those of young stellar objects, microquasars,
gamma-ray bursts,  and AGN jets may be magnetically driven.
Even supernovae may also involve magnetically driven bipolar outflows
(Wheeler, these proc.).  
How MHD jets work and  
where the requisite large scale magnetic fields come from 
are integrated questions, but are usually studied independently.

The large amount of work on MHD jet launching
and collimation will not be reviewed here (see [58,59]).
However, note that the launching mechanisms could be 
divided into ``spring'' [60] and ``fling'' [61] mechanisms. 
In the former class, the jet is launched initially by toroidal magnetic
field pressure. Imagine for example, a dipole magnetic 
field anchored in a star which incurs rapid differential
rotation (such as the collapse from a white dwarf to a young
neutron star as invoked in [62] for gamma-ray bursts) 
or in a supernova core collapse (Wheeler, Meier personal comm.).  
As the differential 
rotation winds up the field, the toroidal field pressure grows quadratically
in time.  When the pressure reaches some critical value, the field
will act something like a coiled spring and can 
drive a strong torsional Alfv\'en wave containing 
directed Poynting flux outward.
This could in principle power a jet. Related mechanisms have
been discussed for disks  [60,63].
In this regard, note that in the case of AGN,
it is not clear if the jet emission we see represents
dissipation from instabilities at the edge of the jet, 
re-acceleration inside the jet along the bulk flow, or just 
emission from a very small number of particles 
carrying the currents [63] which support the magnetic fields.

In the ``fling'' launch mechanism [61],
the initial launch is driven by centrifugal force.
The rigid field lines significantly 
weaken the effective gravitational potential
when sufficiently inclined to the normal, allowing material
fling out along poloidal field lines.
Subsequently, before reaching the Alfv\'en surface,
the driving does become magnetically driven
as in the ``spring'' mechanism.
Since simulation of such launching treats the base of the jet
(e.g. its initial launch point in the corona) 
as a boundary condition, simulators often 
load the field lines with a little mass flow to get the process started
[64].  
The initial 
launching is different for spring and fling, but the ultimate 
collimation mechanisms could be the same, e.g. hoop stresses.
The extent of the collimation appears to be 
sensitive to the boundary conditions of the outflow however [65].

MHD jet luminosity is fueled by the rotational energy.
In systems which have both central compact rotators and 
disks, outflows could emanate from both e.g. [66].  
In black hole systems, the relative contribution to the jet power from 
regions within and outside of the last stable orbit of the disk 
has been addressed [67]. 
Even if a Blandford-Znajek type mechanism is operating from
the hole, a jet from the disk may in fact always
dominate.  However, the black hole spin can still influence the
jet, since it determines the inner edge of the disk, 
and is ultimately important for understanding
the magnetospheres [68].

Where do the required magnetic fields [58] come from?
The first possibility is that they are accreted.
But this may not work for a turbulent accretion disk.
Consider a turbulent disk threaded by a large scale vertical magnetic field.  
The field  is subject to turbulent diffusion and may 
incur a net diffusion outward [69] (with some dependence on turbulent
Prandtl number.) The role of reconnection may
not yet be fully appreciated in this process.
Without reconnection, the mean field is indeed subject to turbulent
diffusion, but it cannot ultimately separate from the gas which has 
a systematic inward motion.  In the absence of a topology change that 
can release field lines from the initial material they thread, 
the field would accrete on the diffusion time scale. 

If the fields cannot be accreted then they
would have to have been threading the central object before the disk
formed, or be generated by a dynamo in the central object 
(if not a black hole) or disk. 
A traditional approach to the amplification of fields on scales
larger than the scale of the input turbulence is the mean field
dynamo [70] mentioned earlier.
Only modes which have an initial seed field
can be amplified.
For an accretion disk dynamo, the limit of field energy density 
is the turbulent energy density.
This limit is $\alpha_{ss}$ times the thermal energy density,
which suggests that ``spring'' mechanisms
are less likely from dynamo produced fields than
``fling'' mechanisms in turbulent disks.  

It is important to distinguish between ``large'' scale and ``mean'' fields.
Standard mean field theory is degenerate with respect
to the topology of the field on scales smaller than the scale of the mean.
Disconnected loops can have the same mean as a connected winding field line.  
In the formalism of  mean field dynamo theory, reconnection
is therefore not strictly required (though it is likely happening anyway).
This is not commonly recognized.  Said another way, neither
the mean field nor the fluctuating component of the field
are the topologically physical field.  To generate jets, common 
wisdom holds that the mean fields actually do have to correspond
to the topologically physical field.  Perhaps this need not be the case
if the Poynting flux driving the jet is an average quantity:
$\lb{\bf E} \times {\bf B}\rb =\lb{\bf e}\times {\bf b}\rb +
{\overline {\bf E}} \times {\overline {\bf B}}$, where the first term
on the right, due to only fluctuating quantities, is usually ignored in this
context (but see [71] where the collimation is non-magnetic.)
If we do ignore  fluctuating components, then to  magnetically 
launch and collimate jets by the ``fling'' mechanism, the mean fields
would need to be topologically large scale.  
One plausible way this could arise in a disk corona is if flux loops
make their way to the surface 
and subsequent reconnection events inverse cascade  
smaller loops to larger loops [40].

The role of boundary conditions is particularly important for mean field dynamos.  First,  generating a net flux of the mean field 
in a quadrupole mode inside an  
object is accomplished by diffusion of the reverse flux through the boundary.
Fast cycles (e.g  solar cycle) of a dipole field  also require 
boundary diffusion to change the flux inside.
Incompressible simulations which employ
periodic boundary conditions over the scale of the mean 
cannot see mean field growth because the induction equation then constrains
the mean field to be time independent. 

In addition, ref. [72] showed that 
conservation of magnetic helicity means that the growth of 
large scale field with one sign of magnetic  helicity inside
the system requires helicity of the opposite sign to diffuse out the
boundary.  Some results showing strong dynamo coefficient
$\alpha$ quenching [73] 
may therefore highligh just an effect from the assumed boundary 
conditions rather than dynamical suppression. In general, to properly 
test large scale field formation in a turbulent disk one must really
utilize a global study, with significant scale separation and 
diffusive boundary conditions.

\subsection*{Closing Comment}

There is much to be learned on all of the above subjects by looking
at the sun [74], as others would also advocate [58].  
Note that dynamos in the Sun may operate
differently than in disks.  For the Sun, strong shear amplification
may take place below the actual turbulent zone, whereas in 
disks, the turbulent zone and the shear are the same region.
However the sun has a large scale wind and with an active corona [74],
consistent with MHD outflows along 
open field lines and x-ray activity resulting
from dissipation of closed field lines.  
We should expect the same for a wide variety of turbulent 
astrophysical rotators.

\bigskip

\noindent {\bf Acknowledgements}: E.B. acknowledges the invitation 
and support of KIAS, and 
support from a DOE Plasma Physics Junior Faculty Development Award.




\begin{references}

\ni  Pringle J.E.,  ARA\&A, {\bf 19}, 137 (1981)

\ni  Papaloizou, J.C.B. \& Lin, D.N.C.,  ARAA, {\bf 33}, 505 (1995);
Rees M.J. ARA\&A, 22, 471 (1984)

\ni  Balbus, S.A. \& Hawley, J.F., 1998, Rev. Mod. Phys, {\bf 70}, 1 (1998)

\ni  Balbus S.A. \& Hawley J.F.,  ApJ, {\bf 376} 214 (1991)

\ni  Igumenshchev I.V.; Abramowicz, M.A. \& Narayan R, ApJ {\bf 537}, L27
(2000)

\ni  Narayan R., Igumenshchev I.V., Abramowicz, M.A. ApJ in press,
astro-ph/9912449 (2000)
 
\ni  Li H., Finn, J.M., Lovelace R.V.E., and 
Colgate S.A., ApJ {\bf 533} 1033 (1999); 
Lovelace R.V.E. et al., ApJ {\bf 513} 805 (1999)

\ni  Narayan R. \& Mahadevan R. \& Quataert E.,  in {\sl Theory of Black Hole Accretion Disks}
ed. M.A. Abramowicz, G. Bjornsson, 
J.E. Pringle Cambridge Univ Press, Cambridge, 1998, p148.

\ni   Ichimaru, S., Ap J, {\bf 214} 840 (1977)

\ni  Rees M.J., in {\it The Galactic Center} ed. G.R. Riegler \& R.D. Blandford AIP, New York, 1980, p. 106; Rees, M.J. et al, Nature {\bf 295}, 17 (1982)

\ni  Narayan R., \& Yi, I., {\bf 428} L13 (1994); 
Narayan R., \& Yi, I., {\bf 452} 710 (1995)

\ni  Di Matteo, T., Blackman E.G. \& Fabian, A.C., MNRAS {\bf 291}, L23  (1997)

\ni  Blandford R.D. \& Begelman M.C., MNRAS {\bf 303}, L1. (1999)

\ni  Quatert E. \& Gruzinov A., in press ApJ, astro-ph/9912440 (2000).

\ni  Gruzinov A., sub to ApJ, astro-ph/9809265 (1998).

\ni  Igumenshchev I.V. \& Abramowicz M.A.  MNRAS {\bf 303}, 309 (1999)

\ni   Brandenburg, A. et al. ApJ {\bf 446} 741. (1995)

\ni   Stone J.M. et al., ApJ {\bf 463} 656 (1995)

\ni   Shakura N.I. \&  Sunyaev R.A.,  A.\&A., {\bf 24}, 337 (1973)

\ni  Balbus S.A., Gammie C.F.,  Hawley J.F.,  MNRAS, {\bf 271}, 197 (1994)

\ni  Blackman, E.G. \& Chou T.C., ApJ, {\bf 489}, L95 (1997)

\ni  Blackman E.G.,  sub. to MNRAS, astro-ph/0006241 (2000)

\ni  Blackman E.G., MNRAS {\bf 299} L48.

\ni  Di Matteo T. et al., MNRAS {\bf 305} 492 (1999)

\ni  Moiseev S.S.et al. Sov. Phys. JETP {\bf 58} 1149 (1983);
Frisch U., Zhe Z.S., Sulem P.L. Physcia D, {\bf 28} 382 (1987); 
Khomenko G.A., Moiseev S.S., Tur A.V., JFM, {\bf 225} 355 (1991); 
Kitchatinov, L. L., 
R{\"u}diger, G., \& Khomenko, G., A.\& A., {\bf 287}, 320 (1994);
Kitchatinov, L. L.,
R{\"u}diger, G., \& Kuker, M., A.\& A., {\bf 292}, 125 (1994)

\ni  Pouquet, A., Frisch, U., \& Leorat, J., JFM {\bf 77}, 321 (1976)

\ni  Kraichnan R.H. \& Montgomery D., Rep. Prog. Phys., {\bf 43} 547 (1980)

\ni  Ingersoll A.P., Science 248, 308 (1990); Marcus P.S.,  ARAA, {\bf 31} 
523 (1993)

\ni  Marcus P.S. JFM, 215 393 (1990); McWilliams, J.C.,  
Phys Fl, {\bf 2} 547 (1990)

\ni  Sommeria J, Meyers S.D., Swinney H.L.,  Nat. {\bf 331} 689 (1988)

\ni  Chavanis P.H. \& Sommeria J, JFM {\bf 356}, 259 (1998)

\ni  Adams F. \& Watkins R., ApJ, {\bf 451} 314 (1995)

\ni  Godon P. \& Livio M., ApJ {\bf 523} 350 (1999)

\ni  Chavanis P.H., A\&A {\bf 356} 1089 (2000)

\ni  Barge P. \& Sommeria J., A\&A, {\bf 295} L1 (1995);
Tanga P. et al.
Icarus, {\bf 121} 158 (1996); 
Hodgson L.S. \& Brandenburg A., A\& A, 
{\bf 330} 1169 (1998);
Bracco A. et al.
Phys. Flu. {\bf 11}, 2280 (1999)

\ni  Politano, H. \& Pouquet, A. Phys Rev E., {\bf 52} 636 (1995)

\ni  Vishniac E., ApJ 446 724 (1995); Vishniac E., ApJ {\bf 451} 816 (1995); 
Blackman E.G. PRL, 77 2694 (1996)

\ni  Galeev, A.A., Rosner, R., Vaiana, G. S., ApJ, {\bf 229} 318 (1979)
 Field, G.B. \& Rogers, R.D., ApJ, {\bf 403} 94 (1993);
 Haardt F. \& Maraschi, L., MNRAS, {\bf 413}, 507 (1993);
Di Matteo T.,  MNRAS 299 L15 (1998); Coroniti F.V.;  
Tout C. \& Pringle J.E.,  MNRAS, {\bf 281} 219 (1996).

\ni  Priest E.R., 1998, Ap\&SS, {\bf 264}, 77.

\ni  Tout C. \& Pringle J.E., MNRAS, {\bf 281} 219 (1996); 
Romanova M.M., 1998, ApJ 500 703 (1998).

\ni  Miller K.A. \& Stone J.M., ApJ {\bf 534} 398 (2000)

\ni  Brandt W.N. et al. MNRAS {\bf 303} L53, (1999).

\ni  Reynolds C.S.,  in  Poutanen, J. \& Svensson R., 
eds., {\sl High Energy Processes in Accreting Black Holes},
ASP Conf. Series Vol. 161, p178 (1999)

\ni  Fabian A.C. Iwasawa K. Reynolds C.S., Young A.J. accepted to PASP,
astro-ph/0004366, (2000).

\ni  Nandra K. et al.
ApJ, {\bf 477}, 602 (1997)

\ni   Tanaka Y. et al., Nature, {\bf 375}, 659 (1995);

\ni  Iwasawa K.
et al.,  MNRAS, {\bf 282}, 1038 (1996); 
Iwasawa K. et al.,  MNRAS, {\bf 306}, L19 (1999);
Lee, J.C. et al., MNRAS, {\bf 310} 973 (1999)


\ni  Pariev V.I. \& Bromley B.C, ApJ, {\bf 508}, 590 (1998)

\ni  Blackman E.G., 1999, MNRAS, {\bf 306}, L25 (1999)

\ni  Hartnoll S.A \& Blackman E.G. MNRAS, in press, astro-ph/9908275 (2000)


\ni  Pringle J.E., 1996, MNRAS, {\bf 281}, 357

\ni   Terquem C. \& Bertout C., A\&A, {\bf 274}, 291 (1995)

\ni  Miyoshi M. et al., Nature, 373, {\bf 127} (1995);
Herrnstein J.R. et al., ApJ, {\bf 468}, L17 (1996)

\ni  Nishiura S., Murayama T. \& Taniguchi Y., PASJ, {\bf 50}, 31 (1998)

\ni  Phinney E.S., in F. Meyer  \emph{et al.}, eds, Theory of Accretion Disks. Kluwer, Dordrecht, p. 457 (1989);
Maloney P.R., in {\sl Highly Redshifted Radio Lines}, 
edited by C.L. Carilli  et al., ASP Conf. Series Vol. 156, 1999, p267

\ni  Turner T.J. et al., ApJ, {\bf 488} 164 (1997)

\ni Guilbert P.W. \& Rees M.J., MNRAS, {\bf 233} 475 (1988); 
 Celotti, A., Fabian, A. C., Rees, M.J. MNRAS {\bf 255} 419 (1992);
Kuncic Z. Blackman E.G., Rees M.J.  MNRAS {\bf 283}, 1322 (1996)

\ni  Blandford R.D., in 
``Proc of Discusison Meeting on Magnetic Activity in Stars, Disks and Quasars.'', Ed. D. Lynden-Bell et al. Phil. Trans. Roy. Soc. A in press (2000);
Ferrari A., ARAA, 36 539 (1998)

\ni   Konigl, A. \&  Pudritz, R. E.,
in {\sl Protostars and Planets IV},
 eds V. Mannings et al., (Tucson: University of Arizona Press); (2000)


\ni  Uchida Y. \& Shibata K., PASJ, 37 31 (1985); 
Contopoulos J.,  ApJ 450 616 (1995); 
Lynden-Bell D., 1996, MNRAS, {\bf 279}, 389 (1996);
Heinz S. \& Begelman M.C., ApJ {\bf 535}, 104 (2000)

\ni  Blandford R.D. \& Payne D.G. MNRAS, {\bf 199}, 883 (1982); 
 Lovelace R.V.E., Wang J.C.L., Sulkanen M.E. ApJ, {\bf 315}, 504 (1987);
Meier D.L., ApJ 522, 753 (1999)

\ni  Ruderman M.A., Tao L., Kluzniak W., ApJ in press, 
astro-ph/0003462 (2000)

\ni  Colgate S.A. \& Li H., ApSS, {\bf 264}, 357 (1999) 

\ni  Krasnopolsky R., Li Z-Y. Blandford R.D., ApJ {\bf 526}, 631 (2000)

\ni  Ustyugova G.V., et al ApJ {\bf 516}, 221 (1999)

\ni  Blackman E.G., Frank. A., Welch C., accepted to ApJ (2000);
astro-ph/0005288.

\ni  Blandford R.D. \& Znajek R.L.  MNRAS, {\bf 179}, 433 (1977);
Punsley B. \& Coroniti F.V., ApJ {\bf 350}, 518 (1990);
Livio M., Ogilvie G.I., Pringle J.E., ApJ {\bf 512}, 100 (1999);

\ni  Khanna R. \& Camenzind M., A\& A, {\bf 307}, 665 (1996);
Krolik J.H.,  ApJ {\bf 515}, 
L73 (1999); Meier D.L., ApJ {\bf 522}, 753 (1999); 
Tomimatsu A., ApJ {\bf 578}, 972 (2000) 



\ni  van Ballegoijen A.A., in {\sl Accretion Disks and Magnetic
Fields in Astrophysics}, Kluwer, Dodrecht 1989, p99; 
Lubow S. H., Papaloizou J. C. B. Pringle J.E., MNRAS, {\bf 267}, 235 (1994)

\ni  Parker, E. N., {\sl Cosmical Magnetic Fields}, Clarendon Press,
Oxford, 1979; R{\"a}dler, K.-H.,
{\sl Generation of Cosmic Magnetic Fields},
Proc. of Mexican School On Astrophysics,  Springer, Heidelberg, 1999, p1.

\ni  Heinz S. \& Begelman M.C., ApJ {\bf 535}, 104 (2000)

\ni  Blackman E.G. \& Field G.B., ApJ {\bf 534}, 597 (2000)

\ni  Cattaneo F., \& Hughes, D.W. Phys. Rev. E. {\bf 54}, 4532 (1996)

\ni Parnell C.E., Ap\&SS, {\bf 261}, 81 (1998);  
Fisk L.A., {\it The Major Discoveries of the ULYSSES Mission}
in 25th International Cosmic Ray Conference, edited by M.S. Potgieter et al., 
World Scientific, River Edge NJ,  1998, p.27.

\end{references}
\end{document}